\documentclass[twocolumn,epjc3]{svjour3}
\smartqed  
\RequirePackage{graphicx}
\graphicspath{{fig/}}
\usepackage{siunitx}
\RequirePackage{booktabs}
\usepackage{rotating}
\usepackage[utf8]{inputenc}
\RequirePackage[numbers,sort&compress]{natbib}
\RequirePackage[colorlinks,citecolor=blue,urlcolor=blue,linkcolor=blue,bookmarksdepth=3,bookmarksopenlevel=3]{hyperref}
\usepackage[british]{babel}
\usepackage{adjustbox}
\usepackage{hhline}
\usepackage{multirow}
\usepackage{hyperref}
\journalname{Eur. Phys. J. C}

\begin{document}

\title{Implication of the Temperature-Dependent Charge Barrier Height of Amorphous Germanium Contact Detector in Searching for Rare Event Physics}

\titlerunning{Amorphous Germanium in Cryogenic Liquids}

\author{R. Panth\thanksref{usd}
\and
W.-Z. Wei\thanksref{usd}
\and
D.-M. Mei\thanksref{usd, e2}
\and
J. Liu\thanksref{usd}
\and
S. Bhattarai\thanksref{usd}
\and
H. Mei\thanksref{usd}
\and
M. Raut\thanksref{usd}
\and
P. Acharya\thanksref{usd}
\and
K. Kooi\thanksref{usd}
\and
G.-J. Wang\thanksref{usd}
}

\thankstext{t1}{This work was supported by NSF OISE-1743790, PHYS-1902577, OIA-1738695, DOE FG02-10ER46709, the Office of Research at the University of South Dakota and a research center supported by the State of South Dakota.}
\thankstext{e2}{Corresponding author E-mail address: \href{mailto:dongming.mei@usd.edu}{dongming.mei@usd.edu}}


\institute{University of South Dakota, 414 East Clark Street, Vermillion, SD 57069, USA\label{usd}
}

\date{Received: date / Accepted: date}

\maketitle
\begin{abstract}
The exploration of germanium (Ge) detectors with amorphous Ge (a-Ge) contacts has drawn attention to the searches for rare-event physics such as dark matter and neutrinoless double-beta decay. The charge barrier height (CBH) of the a-Ge contacts deposited on the detector surface is crucial to suppress the leakage current of the detector in order to achieve la ow-energy detection threshold and high-energy resolution. The temperature-dependent CBH of a-Ge contacts for three Ge detectors is analyzed to study the bulk leakage current (BLC) characteristics. The detectors were fabricated at the University of South Dakota using homegrown crystals. The CBH is determined from the BLC when the detectors are operated in the reverse bias mode with a guard-ring structure, which separates the BLC from the surface leakage current (SLC). The results show that CBH is temperature dependent. The direct relation of the CBH variation to temperature is related to the barrier inhomogeneities created on the interface of a-Ge and crystalline Ge. The inhomogeneities that occur at the interface were analyzed using the Gaussian distribution model for three detectors. The CBH of a-Ge contact is projected to zero temperature. The implication of the CBH at zero temperature is discussed for Ge detectors with a-Ge contacts in searching for rare-event physics.   
\end{abstract}

\section{Introduction}
\label{intro}

Ge detectors with a large electron/hole barrier height are required to obtain a low leakage current, which is thermally generated in the contact materials. Ge detectors are widely used for $\gamma$-ray spectroscopy ~\cite{amman2000position,pehl1977germanium,tavendale1965large,muggleton1972semiconductor}, rare-event physics searches such as neutrinoless double-beta ($0\nu\beta\beta$) decay ~\cite{agostini2020searching,alvis2019search,abgrall2017large} and dark matter ~\cite{aalseth2013cogent,wang2020improved,agnese2018results,mei2018direct}, as well as astroparticle physics ~\cite{aalseth2011astroparticle}, medical imaging ~\cite{hasegawa1991prototype} and homeland security ~\cite{stave2015germanium}.
Ge has a relatively small energy band-gap (0.67 \,eV at room temperature) compared to other semiconductors. Thermal phonons can excite electrons into the conduction band and cause too much noise if Ge detectors are operated at room temperature. Operating Ge detectors at cryogenic temperature reduces thermal noise and hence allows the detectors to perform energy spectroscopy. The low energy band-gap gives the advantage of generating a large number of charge carriers for the equivalent energy deposition from  rare-event physics within a detector and hence increasing its energy resolution. Ge also possesses the advantage of having a high mobility for charge carriers that helps in minimizing the loss of charge carriers from trapping during the transit to the electrode.\\
To make a Ge detector, passivation of a Ge crystal is critical to minimize the surface leakage current (SLC), which is primarily caused by the surface dangling bonds and other surface defects when the surface is exposed to air. Amorphous Ge (a-Ge) is one of the passivation materials used to make rectifying Schottky contacts for Ge detectors. Three advantages make a-Ge contacts attractive for Ge detectors. One is that a-Ge contacts provide bipolar blocking for charge carriers and hence the fabrication process is simpler with only a sputtering machine unlike  traditional Ge detectors made with lithium diffused contacts and boron implanted contacts, which require two different machines. The second advantage is that an a-Ge contact is a layer of a-Ge coated on the surface of Ge detector and hence maximizes the sensitive volume of the detector in contrast to a lithium diffused detector that creates a dead layer as well as a transient layer, which decrease the sensitive volume. Lastly, the contacts can be easily segmented in an a-Ge coated detector as studied by Luke and Amman~\cite{luke1992amorphous, luke1994140, luke2000germanium, amman2000position}. The interface between a-Ge contact and crystalline Ge creates a charge barrier height (CBH), which provides bipolar blocking for the injection of holes or electrons, depending on the sign of the applied bias voltage. The bulk leakage current (BLC) of a detector depends on the CBH of the rectifying contacts. Thus, it is of great importance to study the variation of the CBH as a function of temperature, since the mechanism of charge generation in the a-Ge contact layer is governed by thermionic emission.   

Several papers have been published on the investigation of the CBH of rectifying contacts by studying the current-voltage (I-V) or capacitance-voltage (C-V) characteristics ~\cite{wei2018investigation,wei2020impact,panth2020characterization,mei2020impact,tripathi2012analysis,zhang2019measure,sheoran2020temperature,nouchi2014extraction,di2016tunable}. In this study, we have used the I-V characteristics at different temperatures. We have also implemented a guard-ring structure on the top surface of the detector to separate the BLC from the SLC. The primary source of BLC is the injection of charge carriers from a-Ge contacts to the bulk of the detector. Though, thermal ionization of impurities is also a source of BLC ~\cite{hansen1977amorphous}, it is expected to be much smaller ($\sim$ three orders of magnitude) than the injection from a-Ge contacts when the detector is operating at liquid nitrogen temperature (77 K). Therefore, the measured BLC as a function of bias voltage can be used to study the CBH at different temperatures.\\

\section{Experimental Methods}
\subsection{Fabrication of Ge Detectors}
\begin{figure}[htbp] \centering
  \includegraphics[width=0.9\linewidth]{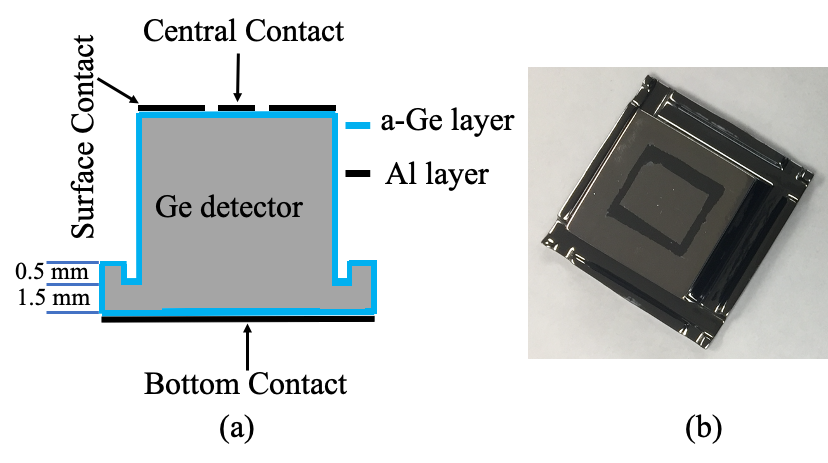}
  \caption{a) Schematic representation of a Ge detector with a guard-ring structure. b) A guard-ring Ge detector fabricated at USD.}
  \label{f:Detector}
\end{figure}

Three guard-ring detectors made from p-type Ge crystals grown at University of South Dakota (USD) were used for the study of CBH. The crystals were grown using the Czochralski method~\cite{wang2015crystal,wang2015high}. High-purity Ge (HPGe) crystals were first sliced into small pieces with a diamond wire saw. Wings and grooves in a size of $\sim$ 2 \,mm each were made on all of the four sides of the HPGe crystal as shown in Figure~\ref{f:Detector}. The purpose of the wings is to allow for handling the HPGe crystals, avoiding direct contact with the sensitive area of the crystal during the fabrication. HPGe crystals were lapped carefully to remove the visible scratches and chips that occur during the cutting process from a diamond saw. Silicon carbide (SiC) and aluminum oxide (Al$_2$O$_3$) powder with 17.5 and 9.5 micron grids, respectively, were used for lapping. The lapped crystals were then etched in a mixture (1:4) of hydrofluoric (HF) and nitric (HNO$_3$) acids to etch away the fine scratches. The etched crystal was then submerged in deionised water and dried with nitrogen gas.
The well-processed HPGe crystals were loaded into the sputtering chamber and a-Ge was deposited on the top and the side surfaces of the crystals. Argon and hydrogen gas mixture (93:7) was used to create the plasma maintaining the chamber pressure of $\sim$14 \,mTorr, which is generated and confined to the space containing a HPGe crystal. Then the crystal was flipped and the same process was duplicated for the bottom side. The thickness of the a-Ge was maintained $\sim$ 600 \,nm for all three detectors.

\begin{figure}[htbp]
  \includegraphics[width=\linewidth]{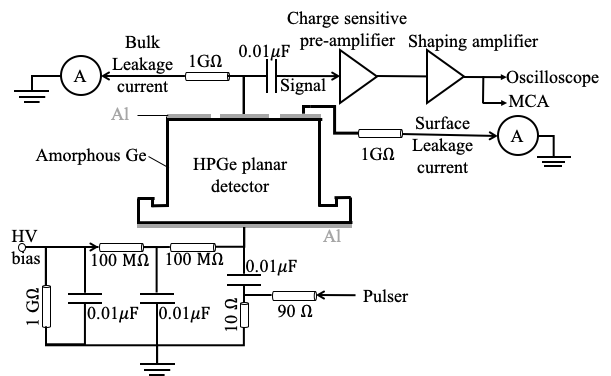}
  \caption{Schematic representation of the electronic circuit for the characterization of a detector.}
  \label{f:circuit}
\end{figure}

Subsequently, a layer of aluminum (Al) with a thickness of $\sim$100 \,nm was deposited on all the sides of the crystal to form a low-resistance contact area using an electron-beam evaporator for the detector USD-R02. For the detector USD-R03 and USD-W03, the Al contacts were sputtered on using the sputtering machine. To sputter Al on the a-Ge surface, plasma was created using argon gas maintaining the chamber pressure of 3 \,mTorr. On the top surface, two contacts were formed by etching out the Al to separate the center contact from the guard-ring (surface) contact. This allows the BLC to be measured through the center contact and the SLC to be measured through the guard-ring contact. A tape mask was used to protect the Al layer. Then the detector was dipped into a 1\% HF solution. The a-Ge was unscathed on all surfaces of the detector. The final contact structures are sketched in Figure~\ref{f:Detector} and \ref{f:circuit}. The full details on the fabrication procedure at USD can be found in the paper published by our group~\cite{meng2019fabrication,Raut:2020eto}.

Note that the completed Ge layers for all three detectors were very uniform after lapping, polishing, and etching, since the same recipe was applied. Therefore, we expect that the completed Ge layer for all three detectors have similar uniformity. This allows us to study how much the inhomogeneous deposition of a-Ge layer during the sputtering process can influence the leakage current.

\subsection{Detector Characterization}
\begin{figure}[htbp] \centering
  \includegraphics[width=0.9\linewidth]{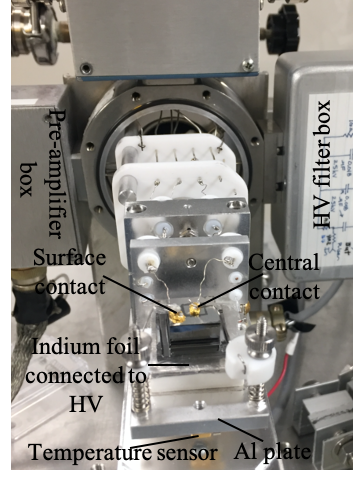}
  \caption{Internal structure of the vacuum cryostat at USD.}
  \label{f:internalsetup}
\end{figure}

The leakage current of each detector was measured in a vacuum cryostat at USD. The schematic representation of the detector characterization set up is shown in  Figures~\ref{f:circuit} and \ref{f:internalsetup}. To monitor the temperature of the detector, a temperature sensor was placed at the bottom of the Al stage. For the detector to be in thermal equilibrium with the Al stage, the measurements were carried out an hour after inserting in the cryostat when the temperature sensor shows the desired temperature.

The leakage current was measured with the combination of a transimpedance amplifier and a multimeter. The voltage signal from the multimeter was converted back to the current signal. The precision of the leakage current measurement from our current setup is 0.1 \,pA.

\begin{figure}[htbp]\centering
  \includegraphics[width=0.9\linewidth]{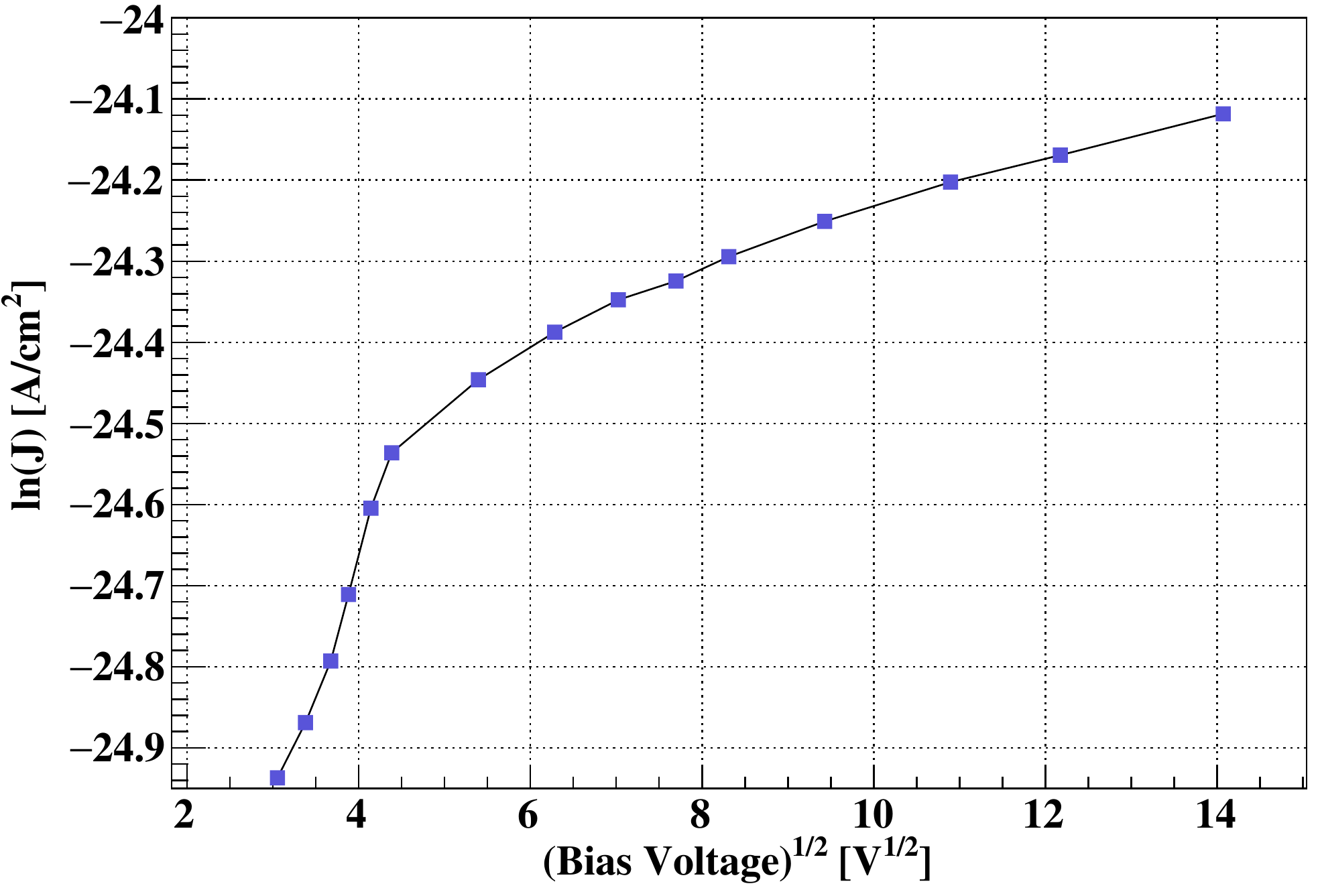}
  \caption{ The variation of the current density versus the square root of bias voltage for USD-R03 detector at 90 \,K. The plot has two different slopes for the voltage range 10-20 and 30-70 \,volts.}
 \label{f:R03-90K}
\end{figure}

The detector with a guard-ring structure allows the characterization of leakage current to be divided into two components: a) BLC and b) SLC. The BLC is mostly dominated by the charge injection from the top and the bottom contacts whereas the SLC is the current created by the surface defects. The variation of the leakage current density versus square root of bias voltage is shown in Figure~\ref{f:R03-90K} as an example. The plot shows that there are two distinguishable regions, which correspond to two different ranges of the applied bias voltage, 10 - 20 volts and 30 - 70 volts. This feature indicates the quantum mechanical properties of the charge carriers. At a lower bias voltage (10 - 20 volts), the reflection coefficient of charge carriers is more pronounced at the boundary than that at a higher bias voltage (30 - 70 volts). For the study of the CBH at different temperatures, we have only taken into account the BLC for the bias voltage in the range of 30-70\,V where thermionic emission dominates. 
\subsection{Charge Blocking Contacts}
Charge carriers are injected through the contacts. To maintain a low leakage current, charge blocking contacts are needed. Lithium-diffused ($n^{+}$) and boron-implanted ($p^{+}$) contacts block holes and electron, respectively. Diffusion of lithium into the detector reduces the sensitive volume of the detector. Since the a-Ge has negligible diffusion into the Ge detector, the sensitive volume of the detector for rare event searches remains intact. Furthermore, the bi-polar blocking behaviour of a-Ge eliminates the need for two different contacts at the top and the bottom of a Ge detector ~\cite{luke1992amorphous}. For a high purity p-type Ge detector passivated with a-Ge contacts and the electrodes formed by deposition of Al, if it is negatively biased from the bottom electrode, then the detector starts to deplete from the top and the BLC is primarily dominated by hole injection from the top contacts. Non-zero conductivity of a-Ge also contributes to the SLC influenced by the hopping conduction mechanism ~\cite{shutt2000solution,bhattarai2020investigation}. The charge collection in a Ge detector is usually carried out with a bias voltage that is a few hundreds volts above the full depletion voltage at which the contribution to the BLC is from both the top and bottom contacts. Analysis of the properties of the charge blocking contacts thoroughly is necessary for the successful operation of the detector. The charge blocking contacts formed by a layer of a-Ge on a Ge crystal are characterized on the basis of: a) the CBH with respect to crystalline Ge; b) the thermal stability; c) the ability of withstanding high bias voltage without breakdown; and  d) the surface inhomogeneity.

\subsection{I-V-T Characteristics and CBH Measurement}

The CBH is calculated based on the current-voltage-temperature (I-V-T) characteristics. The saturation current is defined as the current density corresponding to zero bias voltage and can be obtained by extracting the linear portion of the logarithmic plot of current density versus bias voltage ($V \ge 3kT/q$). The model for the leakage current dependence on temperature and applied bias voltage is developed by D\"ohler, Brodsky~\cite{dohl74,brod75,brodsky75} and Schottky~\cite{sze81}. This model was applied to a-Ge contacts on HPGe detectors as well~\cite{hall}.
Leakage current is directly proportional to temperature as reported in these studies ~\cite{dohl74,brod75,brodsky75,sze81,hall,looker2015leakage,wei2018investigation,wei2020impact,panth2020characterization,mei2020impact}.

The thermionic emission model predicts the current flowing across the metal-semiconductor interface as:
\begin{equation}
J=J_\infty\exp(-\psi_{0,b}/kT)[1-\exp(-qV_d/kT)]f(V_d),
 \label{f:equation1}
\end{equation}
where
$f(V_d)=\exp\{([(2q(V_{bi}+V_d)+N/N_f)N/N_f]^{1/2}-N/N_f)/kT\}$,
$J$ is the ratio of leakage current to the contact area known as leakage current density, $\psi_{0,b}$ is the barrier height at zero bias voltage, $k$ is the Boltzmann constant, $T$ is temperature, $V_{bi}$ is the built-in voltage, $V_d$ is the bias voltage, $N$ is the net impurity concentration, $N_f$ is the density of localized energy states near the Fermi level.
$J_\infty$ equals to $A^*T^2$ in the case of a metal contact made on crystalline semiconductor and $A^*$ is the Richardson constant. Since electric field penetration through the contacts is negligible when the bias voltage is low, the value of $f(V_d)$ is usually close to 1. It is worth mentioning that $J_\infty$ can be replaced by $J_\infty=J_0 T^2$~\cite{amman2018optimization}. If the $V_d$ $\>>>$ $V_{bi}$, or $kT/q$, or $N/qN_f$ equation~\ref{f:equation1} reduces to
\begin{equation}
J=J_0 T^2\exp(-\psi_{0,b}/kT)\exp[(2qV_dN/N_f)^{1/2}/kT]
    \label{f:equation2}
\end{equation}
for a partially-depleted detector.
$\Delta\psi = \sqrt{2qV_dN/N_f}$
is the barrier lowering term. $\Delta\psi$ is directly proportional to the applied bias voltage, impurity concentration, and inversely to the density of defect states near the Fermi level. Therefore, to keep the barrier lowering value at a minimum, the density of localized energy states near the Fermi level should be high and the impurity concentration should be low. $J_0$ is constant, the pre-factor, which is left as an open parameter to be determined from the measurements \cite{amman2018optimization}. To calculate $J_0$ from equation~\ref{f:equation2}, the barrier height should be treated as constant with respect to temperature. However, several pioneers have clearly demonstrated that the barrier height is not constant with temperature~\cite{zeyrek2008double,toumi2009gaussian,sil2020elucidation,nicholls2019description,huang2013barrier,zeghdar2015inhomogeneous,karboyan2013analysis}. 
Such a phenomenon indicates there is a barrier inhomogeneity at the interface of a-Ge and crystalline Ge. 

In order to study the variation of CBH as a function of temperature in the a-Ge contacts created at USD, we have treated the pre-factor as constant and equal to the Richardson constant (48 \,A/cm$^{-2}$K$^{-2})$ in the case of the a-Ge deposition on the p-type crystalline Ge~\cite{wei2020impact} and have determined the barrier height at different temperatures.

\begin{figure}[htbp]\centering
  \includegraphics[width=\linewidth]{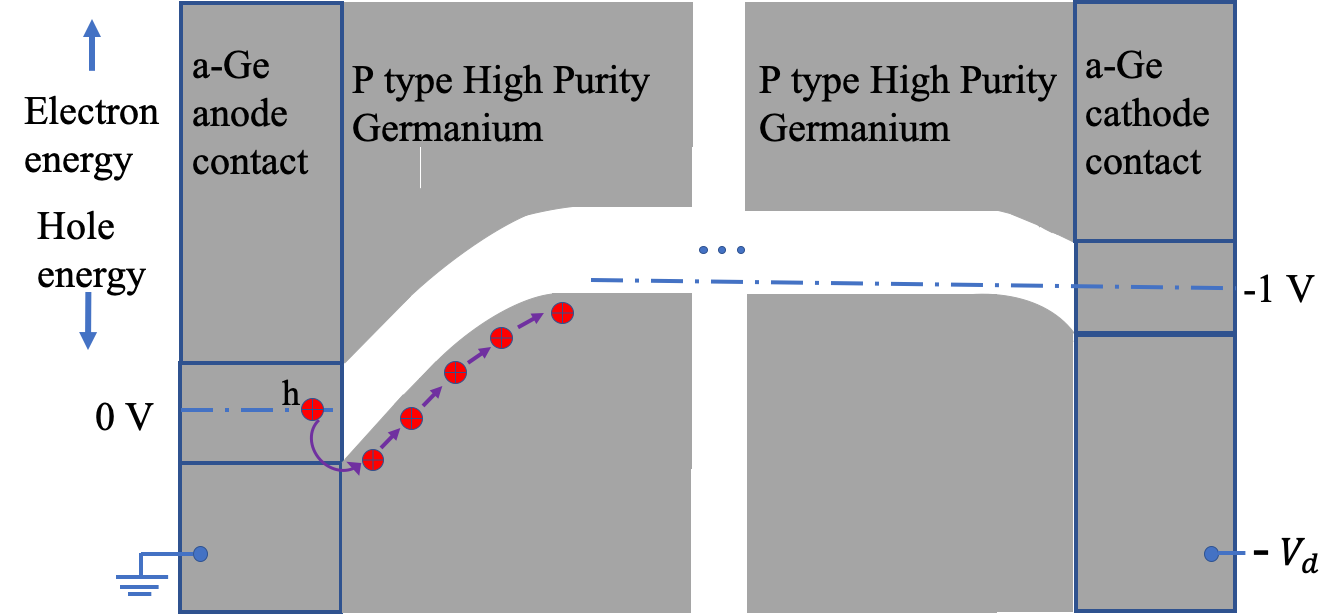}
  \caption{Schematic representation of energy diagram when a small reverse bias is applied to the bottom contact of the p-type Ge detector~\cite{amman2018optimization}. (not to scale)}
  \label{f:EnergyDiagram}
\end{figure}

Utilizing J$_0$ equals A$^*$T$^2$,  equation~\ref{f:equation2} can be re-written as 
\begin{equation}
J = A^*T^2\exp(-\psi_{0,b}/kT)\exp[(2qV_dN/ N_f)^{1/2}/kT].
    \label{f:equation3}
\end{equation}
The Y-intercept obtained from the plot of lnJ versus the square root of the reverse bias voltage gives the zero-bias current density and is given by
\begin{equation}
J_0=A^*T^2\exp-(\psi_{0,b}/kT).
 \label{f:equation4}
\end{equation}
The zero-bias barrier height can be expressed as\\ 
\begin{equation}
\psi_{0,b}=kTln(J_0/A^*T^2).
\end{equation}
Since A$^*$ = 48 \,A/cm$^{-2}$K$^{-2}$ is a constant, thus, we can calculate  $\psi_{0,b}$ for a given temperature using equation~\ref{f:equation4}. Figure~\ref{f:EnergyDiagram} exhibits the energy band diagram for an interface between a layer of a-Ge and a p-type Ge crystalline structure. The variation of the zero-bias barrier height as a function of temperature indicates the inhomogeneity of the coated a-Ge layer. 

It is worth pointing out that the slope obtained from the plot of lnJ versus the square root of the reverse bias gives the density of localized energy states near the Fermi energy level for the a-Ge layer sputtered onto a-Ge surface for a given detector at a given temperature. The relationship between N$_{f}$ and temperature can also be used to study the properties of the a-Ge layer, for example, the resistance or conductivity. However, it does not directly relate to the zero-bias barrier height, as described in equation~\ref{f:equation3}. It may contribute to the systematic errors of the zero-bias barrier height in an indirect way, which is complicated and beyond the scope of this paper. 

\begin{figure}[htbp]\centering
  \includegraphics[width=\linewidth]{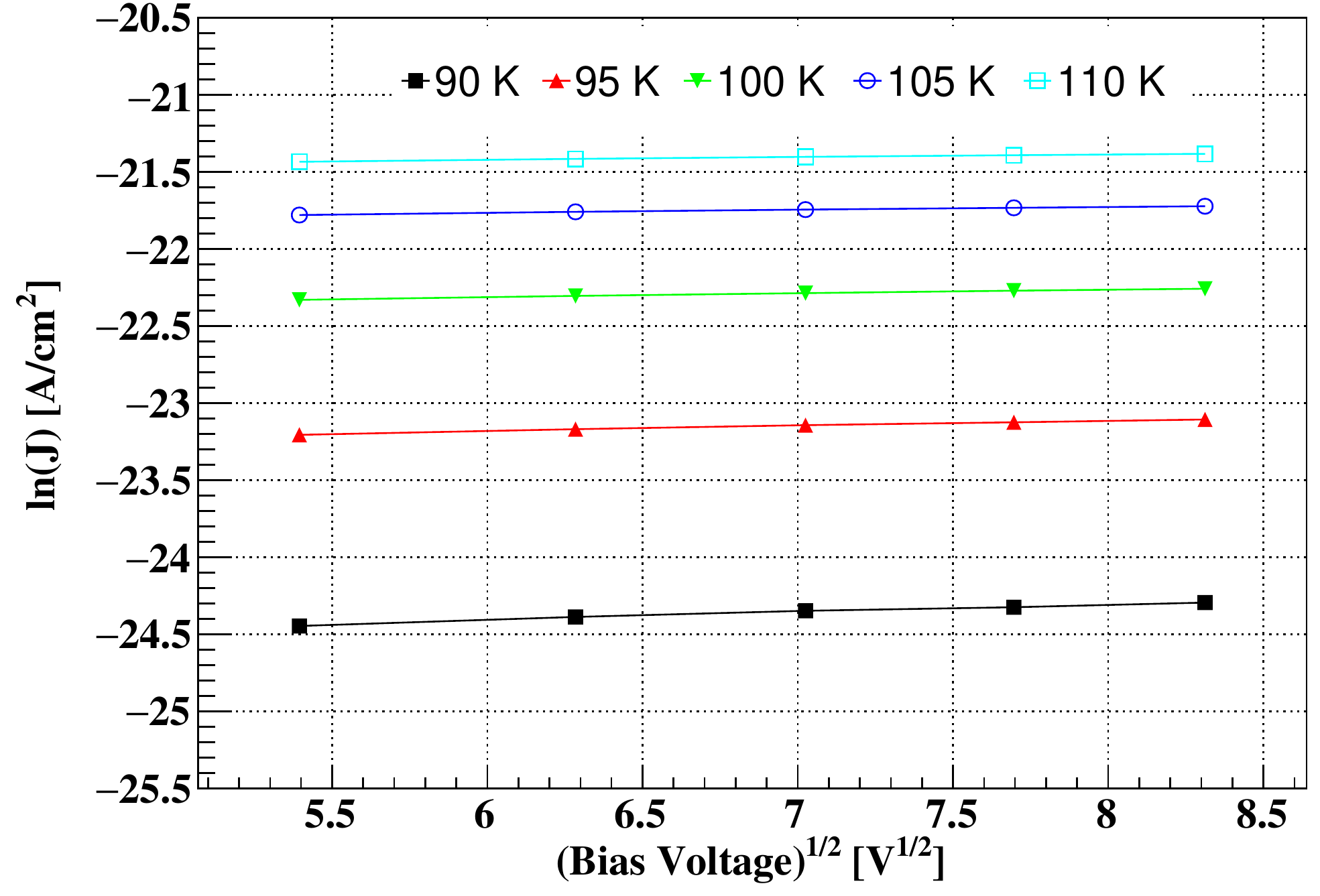}
  \caption{ The variation of the leakage current density versus the square root of bias voltage for USD-R03 detector at different temperatures.}
  \label{f:R03-30-70V}
\end{figure}
\begin{figure}[htbp]\centering
  \includegraphics[width=\linewidth]{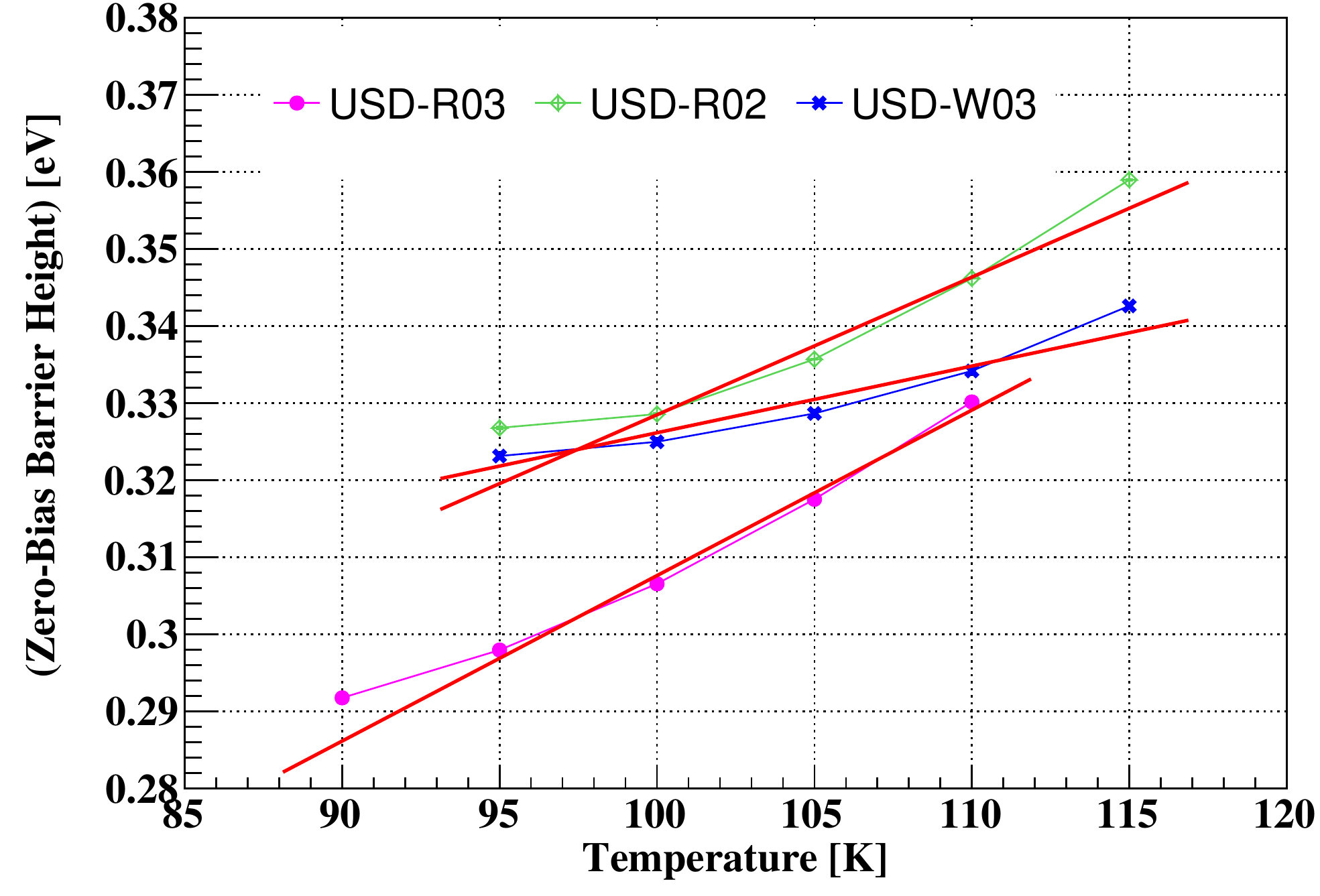}
  \caption{ The variation of the zero-bias barrier height versus temperature for three detectors. The CBH at 0 \,K was extracted from the polynomial fit.}
  \label{f:BHE}
\end{figure}

As an example, the variation of the BLC density versus the square root of bias voltage for USD-R03 detector is shown in Figure~\ref{f:R03-30-70V} at the temperatures 90 \,K, 95 \,K, 100 \,K, 105 \,K and 110 \,K. Similar I-V-T characteristics were obtained for USD-R02 and USD-W03 detectors. The $Y-$intercept obtained from the plots gives the zero-bias BLC density which was used to calculate the zero-bias barrier height. Figure~\ref{f:BHE} displays the dependence of calculated barrier height versus temperature for three detectors. When temperature increases, the barrier height increases or vice versa. 

The zero-bias barrier height is extracted at 0 \,K using a linear fit to the data points of zero-bias barrier height versus temperature. However, there exists uncertainty in the extrapolated barrier height because the zero-bias barrier height versus temperature is not perfectly linear as shown in Figure~\ref{f:BHE}. The intercept of the plot gives the barrier height at zero temperature. The slope in Figure~\ref{f:BHE} represents the variation of the barrier height at a given temperature. The extrapolated zero-bias barrier height at 0 \,K and the slope obtained from the plot ~\ref{f:BHE} is shown in Table ~\ref{t:det1}.
We explore the origin of this variation below. 

\begin{table}[htbp]
  \caption{Barrier height extrapolated from a linear fit of zero-bias barrier height versus temperature.}\label{t:det1}
  \begin{tabular}{lccc}\toprule
   Detector & $^\vdash$ $\psi_{0,b}/eV$ & $^\ast$ $slope$& \\\midrule
   
    $  USD-R02$ & $0.14983\pm$ $8.4E-4$ & $ 0.00178 \pm $ $7.8E-6$ &\\
   $ USD-R03$  &$0.09285\pm$ $5.8E-4$ & $0.00214$ $\pm$ $5.6E-6$ & \\
    $  USD-W03$ & $0.23962\pm$ $1.1E-3$ & $ 0.00086\pm$ $1.1E-5$&\\
   \bottomrule
\end{tabular}

$^\vdash$ The zero-bias barrier height extrapolated to 0 $K$ .\\
$^\ast$ The slope obtained from barrier height versus temperature plot.\\
\end{table}

Note that there is a tendency that the zero-bias barrier height seem to saturate to a value above 0.32 eV for USD-R02 and USD-W03. The zero-bias barrier height seems to saturate to a value above 0.29 eV for USD-R03. To predict the zero-bias barrier height at 0\,K, we take a conservative approach by assuming that the zero-bias barrier height varies linearly as a function of temperature. This is the worst case scenario according to the data points shown in Figure~\ref{f:BHE}, since it will predict the least value for the zero-bias barrier height at 0\,K and hence estimating a maximum value for the bulk leakage current. However, it makes our prediction on the safest side in terms of the bulk leakage current.    

\subsection{The Relation Between the Inhomogeneity of a-Ge Layer and CBH}
A homogeneous interface layer allows us to predict the BLC at different temperature and applied bias voltage using equation~\ref{f:equation2}, since the barrier height is a constant. However, for an inhomogeneous interface layer, the barrier height cannot be treated as a constant value with respect to temperature.
The analysis of the I-V-T characteristics for three different detectors shows that the variation of CBH with respect to temperature is governed by the inhomogeneities at the interface of a-Ge and crystalline Ge. The existence of inhomogeneity at the interface might be related to the cleanliness of the surface of crystalline Ge, the vacuum level inside the sputtering chamber, the stability of the gas flow while creating the plasma, and the variation in the thickness of a-Ge layer.\\
The vacuum pressure of the chamber for all three detectors was maintained less than $4\times 10^{-6}$ \,Torr. However, the gas flow rate was not stable. The pressure of the argon and hydrogen gas mixture in the chamber jumps between 12-16 \,mTorr. The instability of the chamber pressure during the plasma formation caused the unstable condition for the reflected power in the radio-frequency sputtering machine. Exposure to air during fabrication of the detector will be monitored in future detector fabrications to study its role in the nature of Schottky barrier. \\
The Gaussian distribution model developed by Werner and Guttler was applied to explain the correlation between the barrier height variation and the inhomogeneities of the interface~\cite{werner1991barrier}. The expression of the Gaussian model is described as: 
\begin{equation}
\psi_{0,b}=\overline{\psi}-\sigma^2/2kT,
 \label{f:equation6}
\end{equation}
\begin{figure}[htbp]\centering
  \includegraphics[width=\linewidth]{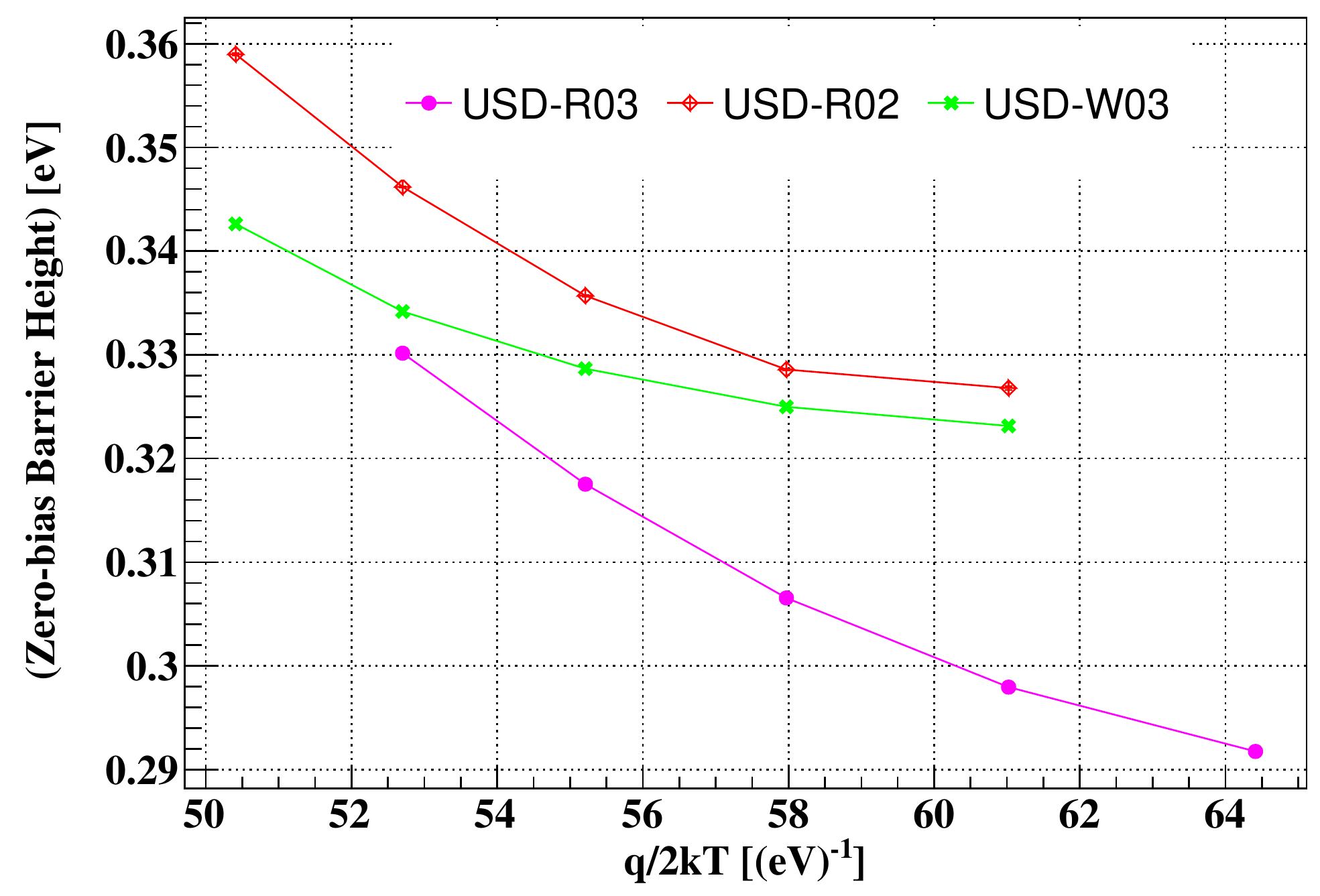}
  \caption{Variation of zero-bias barrier height with $q/2kT$ for three detectors.}
  \label{f:meanbheight30-70V}
\end{figure}
Table~\ref{t:det2} summarizes the properties of the a-Ge contacts deposited on the high-purity Ge crystal grown at USD.

\begin{table*}[htbp]
\begin{center}
  \caption{Summary of three USD detector properties}\label{t:det2}
  \begin{tabular}{lccc}\toprule
    Detector & USD-R02 & $ USD-R03$ & $ USD-W03 $ \\\midrule
    $^\triangleright$ Impurity/cm$^3$ & $2.93\times10^{10}$ & $3.78\times10^{10}$ & $2.60\times10^{10}$ \\
    Thickness/cm & $0.65$ & $0.81$ &$ 0.94$ \\
    $^\ddagger$ Area/cm$^2$ & $ 0.29 $ & $0.48$ & $0.24$\\
    $^\star$ $V_{fd}$/V & $700$ & $1400$ & $1300$ \\
    $^\vdash$ $\psi_{0,b}/eV $@$ 90 K$ & $-$ & $ 0.29174$$\pm$ $1.8E-4$ & $-$\\
    $^\vdash$ $\psi_{0,b}/eV $@$ 95 K$ & $0.32679$$\pm$ $1.8E-4$ & $ 0.29795 $$\pm$ $9.7E-5$ & $0.32313$$\pm$ $1.3E-4$\\
    $^\vdash$ $\psi_{0,b}/eV $@$ 100 K$ & $0.32858$$\pm$ $9.3E-5$ & $ 0.30655 $$\pm$ $5.4E-5$ & $0.32498$$\pm$ $1.5E-4$\\
    $^\vdash$ $\psi_{0,b}/eV $@$ 105 K$ & $0.33570$ $\pm$ $6.4E-5$& $ 0.31752 $$\pm$ $5.4E-5$ & $0.32866$$\pm$ $1.6E-4$\\
    $^\vdash$ $\psi_{0,b}/eV $@$ 110 K$ & $0.34619$$\pm$ $6.7E-5$ & $ 0.33015 $$\pm$ $5.3E-5$ & $0.33418$$\pm$ $1.8E-4$\\
    $^\vdash$ $\psi_{0,b}/eV $@$ 115 K$ & $0.35898$$\pm$ $1.2E-4$ & $ - $ & $0.34262$$\pm$ $2.4E-4$\\
    $^\ast$ $\overline{\psi}$/eV & $0.52367$  $\pm$ $8.2E-4$ & $0.52359$ $\pm$ $8.3E-4$ & $ 0.41734$ $\pm$ $1.1E-3$ \\
    $^\pm $ $\sigma^2$/(eV)$^2$ & $0.00336$ $\pm$ $1.5E-5$ & $0.00371$ $\pm$ $1.4E-5$ & $0.00156$ $\pm$ $2.7E-5$ \\
   \bottomrule
\end{tabular}
\end{center}
$^\triangleright$ Net impurity concentration calculated from the C-V measurements.\\
$^\ddagger$ Area of the central contact on the top surface.\\
$^\star$ Full depletion voltage for the detector.\\
$^\vdash$ Zero-bias barrier height.\\
$^\ast$ Mean-barrier height.\\
$^\pm$ Variance of barrier height fluctuation.\\
\end{table*}
where $\overline{\psi}$ is the mean barrier height and $\sigma$ is the standard deviation. $\sigma$ is assumed to be a constant with respect to temperature for this calculation. The Y-intercept of the plot in Figure~\ref{f:meanbheight30-70V} gives the value of $\overline{\psi}$ and the slope determines the value of $\sigma^2$, respectively. The values of $\overline{\psi}$ and $\sigma^2$ for three detectors are shown in Table~\ref{t:det2}. As can be seen from Table~\ref{t:det2}, the value of $\sigma$ is smaller for USD-W03 detector, which indicates the barrier height fluctuation is smaller than the other two detectors. This implies that the variation of the BLC from USD-W03 will be less than the other two detectors when increasing or decreasing temperature. The standard deviation measured from the Gaussian distribution of barrier height indicates that the barrier inhomogeneity can not be neglected while calculating the barrier height. The deviation of $\sigma$ with respect to $\overline{\psi}$ is within the range of $9 \% - 12 \%$  for all the detectors that were used for this study. 

It is worth mentioning that the Gaussian model is used to study the barrier inhomogeneity. A smaller value of $\sigma$ indicates a more homogeneous barrier and hence pointing to a better sputtering process in the fabrication of a-Ge contacts. 
Since the value of $\sigma^2$ indicates the inhomogeneity of CBH, which is related to the fabrication process, we expect to improve the fabrication process to keep the value of $\sigma^2$ as small as possible.  

\section{Conclusions}
The nature of the Schottky barrier is influenced by the surface properties of the substrate in which the contacts are formed. The barrier height of a-Ge contacts deposited on Ge crystal is calculated using I-V-T characteristics. 
The variation of the Schottky barrier height at different temperatures has been explained by considering the Gaussian distribution model. The inhomogeneity of a-Ge contacts created on the crystalline surface is the main source of barrier height fluctuation with respect to temperature. The observed inhomogeneity difference in the a-Ge layers from three USD detectors suggests that the fabrication process can be improved to obtain a smaller variation in the barrier height for rare-event physics searches. 
Although the results obtained in this study are from a narrow range (90 \,K to 115 \,K), a conservative extrapolation using a linear function predicts a minimum value of zero-bias barrier height at 0\,K. This extrapolated minimum barrier height at 0 K indicates that the injection leakage current must be extremely small owing to the CBH of $\sim$ 0.1 \,eV, which is sufficient to block electrons or holes at low temperature. The Conservative extrapolation of CBH for 3 detectors is shown in Table ~\ref{t:det1}. Using equation~\ref{f:equation4}, a small CBH of 0.02\,eV is enough to block the injection of charge carriers to a negligible level ($\sim 10^{-24}$ A) at helium temperature, suitable for rare-event physics searches. From Table~~\ref{t:det1}, the least values of CBH for three detectors are all significantly larger than 0.02\, eV. Therefore, we conclude that the Ge detectors made with a-Ge contacts are suitable for rare-event physics searches. It is clear that further measurement near 4 K can be valuable to verify the prediction.
\begin{acknowledgement}
  The authors would like to thank Mark Amman for his instruction on fabricating planar detectors, and the Nuclear Science Division at Lawrence Berkeley National Laboratory for providing the vacuum cryostat. We would also like to thank Christina Keller for a careful reading of this manuscript. 
\end{acknowledgement}

\bibliography{refs.bib}
\bibliographystyle{spphys}

\end{document}